\documentstyle[
frontieres
]{article} 
\begin{document} 
\newcommand{\gw}{gravitational wave}
\newcommand{\grad}{gravitational radiation}
\newcommand{\dtc}{detector}
\newcommand{\lsr}{laser interferometer}
\newcommand{\lsri}{laser interferometric}
\newcommand{\sn}{supernova}
\newcommand{\cb}{coalescing binary}
\newcommand{\cbs}{coalescing binaries}
\newcommand{\ns}{neutron star}
\newcommand{\bh}{black hole}
\newcommand{\cw}{continuous-wave source}
\newcommand{\stc}{stochastic background}
\newcommand{\frq}{frequency}
\newcommand{\frqs}{frequencies}
\newcommand{\sens}{sensitivity}
\newcommand{\gcoll}{gravitational collapse}
\newcommand{\grel}{general relativity}
\newcommand{\whdw}{white dwarf}
\newcommand{\msolar}{M_{\odot}}
\newcommand{\pwrten}[1]{10^{#1}}
\newcommand{\mptr}{{\cal M}}
\newcommand{\spec}{S_{h}(f)}
\newcommand{\fracparen}[2]{\left(\frac{#1}{#2}\right)}
\newcommand{\fracsqb}[2]{\left\lbrack\frac{#1}{#2}\right\rbrack}
\newcommand{\ibarddd}{\mbox{$\stackrel{\ldots}{\mbox{\ibar}}$}}
\newcommand{\ibar}{\mbox{\rlap{$I$}--}}
\newcommand{\sups}{{}^}
\newcommand{\subs}{{}_}
\newcommand{\oderiv}[2]{\frac{d #1}{d #2}}
\newcommand{\oderivn}[3]{\frac{d^{#3}\!#1}{d #2^{#3}}}
\newcommand{\pderiv}[2]{\frac{\partial #1}{\partial #2}}
\newcommand{\pderivn}[3]{\frac{\partial^{#3}\!#1}{\partial #2^{#3}}}
\newcommand{\delsq}{\nabla^{2}}
\newcommand{\twopii}{2\pi\imath}
\newcommand{\oderivf}[2]{{d #1}\!/{d #2}}
\newcommand{\oderivnf}[3]{{d^{#3}\!#1}\!/{d #2^{#3}}}
\newcommand{\pderivf}[2]{{\partial #1}\!/{\partial #2}}
\newcommand{\pderivnf}[3]{{\partial^{#3}\!#1}\!/{\partial #2^{#3}}}
\newcommand{\setof}[4]{\{$#1_{#2},\; #2=#3,\ldots,#4$\}}
\newcommand{\tensortype}[2]{{#1}\choose{#2}}
\newcommand{\E}[1]{\times 10^{#1}}
\newcommand{\snr}{signal-to-noise ratio}
\newcommand{\requirements}[1]{\vspace*{10mm}{\centering\framebox[120mm]
  {\begin{minipage}{115mm}#1\end{minipage}}}\vspace*{10mm}}
\newcommand{\splA}[2]{#2}
\newcommand{\splB}[2]{#1}
\newcommand{\American}{\let\spl=\splA} 
\newcommand{\British}{\let\spl=\splB} 
\newcommand{\re}{\spl{re}{er}}
\newcommand{\ou}{\spl{ou}{o}} 
\newcommand{\se}{\spl{se}{ze}}
\newcommand{\aE}{\spl{ae}{e}}
\newcommand{\Ae}{\spl{ae}{a}}
\newcommand{\s}{\spl{s}{z}}
\newcommand{\gindex}[1]{#1\index{#1}} 
\newcommand{\Eqref}[1]{Eq.~(\ref{#1})}
\newcommand{\Figref}[1]{Figure~\ref{#1}}
\newcommand{\Chref}[1]{Chapter~\ref{#1}}
\newcommand{\Apref}[1]{Appendix~\ref{#1}}
\newcommand{\Tabref}[1]{Table~\ref{#1}}
\newcommand{\Secref}[1]{Section~(\ref{#1})}
\newcommand{\hide}[1]{}\newcommand{\units}{\rm\;}
\newcommand{\Ggw}{the University of Glasgow}
\newcommand{\MPQ}{the Max Planck Institute for Quantum Optics}
\newcommand{\BFS}{Bernard F. Schutz}
\newcommand{\Cardiff}{Department of Physics and Astronomy\\University of Wales College of Cardiff, Cardiff, U.K.}
\newcommand{\AEI}{Max Planck Institute for Gravitational Physics\\The Albert Einstein Institute\\Potsdam, Germany}
\heading{%
%
Sources of radiation from neutron stars
%
}  
\par\medskip\noindent 
\author{
\BFS$^{1,2}$ 
} 
\address{
Max Planck Institute for Gravitational Physics (The Albert Einstein Institute), Potsdam, Germany 
} 
\address{
Department of Physics and Astronomy, University of Wales College of Cardiff, Cardiff, U.K.
} 
\begin{abstract} 
I give a brief introduction to the problem of detecting gravitational radiation from neutron stars.  After a review of the mechanisms by which such stars may produce radiation, I consider the different search strategies appropriate to the different kinds of sources: isolated known pulsars, neutron stars in binaries, and unseen neutron stars.  The problem of an all-sky survey for unseen stars is the most taxing one that we face in analysing data from interferometers.  I describe the kinds of hierarchical methods that are now being investigated to reach the maximal sensitivity, and I suggest a replacement for standard Fourier-transform search methods that requires fewer floating-point operations for Fourier-based searches over large parameter spaces, and in addition is highly parallelizable, working just as well on a loosely coupled network of workstations as on a tightly coupled parallel computer.
\end{abstract}
\section{Introduction}
The interferometers now under construction will operate initially with a sensitivity that does not guarantee that they will see any sources, and almost certainly what they do see will be difficult to dig out of the noise.   They can therefore benefit greatly from a careful study of the expected characteristics of potential sources and waveforms.  Probably the most well-studied and understood source is the coalescing compact-object binary, and it has received much attention at this meeting.  Indeed, black-hole binaries may be among the first detections by the new instruments (see talk by Flanagan at this meeting), or they may not exist in sufficient numbers to be seen for many years (see talk by Yungelson at this meeting).  

Less attention has been paid to detecting continuous radiation from neutron stars, although attention is increasing \cite{Brady}.  These could be particularly important in the first years, when the GEO600 detector (see the talk by Ward at this meeting) may be operating on its own and will therefore not be able to identify short bursts of radiation from coalescing binaries reliably.  At the Albert Einstein Institute (AEI) in Potsdam we have initiated a concentrated effort both to understand the relevant physics of neutron stars and to devise efficient algorithms for the signal analysis.  Bar detectors have essentially the same data-analysis problem as interferometers do when looking for this kind of radiation (see the talk by Johnson et al at this meeting), since the signal bandwidth is much smaller than a bar detector's observing bandwidth.  For that reason we have begun a collaboration with the Rome group to study this problem together.  We also are continuing a collaboration with the Caltech group associated with LIGO.

A number of talks at this meeting stem from this effort.  Brady (Caltech) will speak about the data-analysis technique he calls {\em stacking}, where Fourier transforms of data covering relatively short observation periods are incoherently combined by adding their power spectra.  This is an important method for improving the depth of our data searches.  Papa (AEI and Rome) will introduce what seems to me to be a viable alternative to stacking (and the only viable alternative that I know of at present), which I call {\em tracking}, where weak lines in individual Fourier transforms are followed in successive data sets in an attempt to identify persistent signals.  Krolak (AEI and Warsaw) will consider the problem of coherently filtering long stretches of data for signals from pulsars.  Sintes (AEI) will describe a new and exciting technique for removing troublesome interference from external coherent signals (such as the electricity supply frequency and its harmonics), which can obscure wide frequency bands.  Her approach can leave the spectrum clean enough to see a true gravitational wave that had been buried in the interference.  

In the present talk, I will briefly survey the physics of neutron stars as gravitational wave sources, highlighting what seem to me the unanswered questions that are most important for predictions of the strength of these sources.  Then I will turn to the analysis of long stretches of data, and describe a hierarchy of problems of increasing difficulty: looking for sources whose position and frequency are known, searching for sources whose position is known but which may be in a hidden binary system, and finally surveying the entire sky for sources that are not hinted at by any other astronomical observation.  For this latter problem I show the important role that stacking/tracking will play, and describe the difference in sensitivity of searches with and without these techniques.  Finally I suggest what seems to me to be an efficient method of performing the short-period coherent searches that form the input to the stack/track step.  

My conclusion regarding a full-sky survey is that it is daunting but by no means hopeless.  If we have access to a teraflop-class computer -- the fastest we can hope to use in the near future --  then a search of the sky using the full sensitivity that the detectors can achieve over several months seems possible.   Of course, access to such a computer is not guaranteed at present, but without it the sensitivity of these first detectors will inevitably be reduced.

\section{Physics of neutron star gravitational wave sources}

The following is a list of possible mechanisms whereby a neutron star may emit long-duration trains of gravitational waves.
\begin{itemize}
\item {\bf Lumpy stars}  Stars may be deformed in a non-axisymmetric way.  Current theory suggests the crust of a neutron star cannot sustain an ellipticity greater than $10^{-5}$.  This would put radiation from most pulsars (except the Crab) too low for the first generation of detectors.  Bonazzola (see his talk at this meeting) has suggested that buried magnetic fields may sustain a larger ellipticity, but the inhomogeneities of such fields need to be extreme.
\item {\bf Wobbly stars}  If the spin and angular momentum are not aligned, a star will undergo free precession to the extent that it behaves like a solid body.  Its solid crust, and some rigidity given by its magnetic field, probably allow this behavior.  The natural decay time to gravitational radiation is short for fast pulsars, but a certain amount of wobble might be generated from electromagnetic radiation reaction torques created by the star's strong magnetic field (private communication from D.\ I.\ Jones, following an idea due to Goldreich).
\item {\bf Normal modes}  After formation of a neutron star, the normal modes emit radiation as the residual non-axisymmetry decays, but this can be at frequencies too high for interferometers and present-day bars, and it decays rapidly.  Two families of modes may be unstable: $f$-modes (discussed by Chandrasekhar, Friedman, and Schutz \cite{friedman_cfs}) and $r$-modes (as recently suggested by Andersson \cite{Andersson}).  For the f-mode instability, a star has to spin faster than a critical onset spin.  Once there, the mode will radiate away the excess spin, but this could take millions of years for some modes.  For $r$-modes, the star is unstable at any spin, but the growth time of the instability is short only for rapid spins.  Both instabilities may be eliminated if viscosity is high enough, so it may be that only young neutron stars will experience these instabilities. \cite{lindblom}
\item {\bf Accreting stars}  It seems to me that one of the most promising ways of generating gravitational waves from neutron stars is for accretion to excite non-axisymmetries.  Wagoner \cite{wagoner} long ago pointed out that accretion could drive the unstable f-modes (and by implication also the unstable r-modes) into strong radiation.  Or it could increase the wobble angle if the magnetic field of the star acts as a kind of ``gate'' for the accretion (D.\ I.\ Jones, private communication).  Such stars may be in conventional X-ray binaries, or they may be hidden inside O/B stars during the phase of evolution of an X-ray binary in which the neutron star is enveloped in the atmosphere of its companion.  Accretion rates for known X-ray pulsars go up to 0.1 Eddington, but in principle they could be orders of magnitude higher (say, in O/B stars), and the gravitational radiation emitted would be proportional to the accretion rate in such mechanisms.  New X-ray observations may shed light on this, but the common-envelope systems emit no X-rays and are hard to distinguish from ordinary giants.
\end{itemize}

From this list it is clear that there are promising emission mechanisms for neutron stars, but none that are in any degree certain.  It may be that only gravitational wave observations will tell us whether some or all of these mechanisms operate.  Since lumpy stars may be dead as pulsars, and since common-envelope neutron stars may be hidden from electromagnetic observations, only extensive and deep surveys of the sky will answer these questions.

\section{Directed searches: looking for radiation from known pulsars}
This is the simplest kind of continuous gravitational wave search, since one knows in principle both the position and frequency of the pulsar.  But there are complications, since the accuracy with which we have to know the signal for a one-year observation is considerable: the waveform template for a matched filter should contain second and even third time-derivatives of the pulsar's frequency, the pulsar's proper motion may be a factor, and the orientation of the spin axis will not usually be known.  
\begin{itemize}
\item The {\bf position} of a pulsar is known from radio observations, the {\bf pulsar frequency} 
$f_{psr}$ is known, and the {\bf spindown rate} $df_{psr}/dt$ is (sometimes) known.  Data sampled at the detector can be reduced to the solar-system barycenter by stroboscopic sampling \cite{SchutzBlair} or other methods.  The FFT would take only a few seconds on a 1~Tflop computer for a 1-year data set.  But it will still have to be processed as follows to ensure detection.
\item The rotation of the Earth induces an amplitude modulation that is not removed by sampling the phase at the barycenter.  It splits the signal into sidebands that need to be combined, but the way to do this depends on the signal's polarization, hence on the orientation of the pulsar's spin axis (see Krolak at this meeting).
\item The gravitational wave frequency $f_{gw}$ is not necessarily an exact integer multiple $f_{psr}$.  If there is free precession, the gravitational wave frequency could differ by the precession rate:
\[ f_{gw} = 2f_{psr}\left[ 1 \pm \left(\frac{f_{psr}}{1.5\;\rm kHz}\right)^2\right].\]
For the Crab this is $\pm 24$~mHz,  or $7\times 10^5$ frequency resolution bins in a 1-year observation.  An FFT will search this band automatically, but it also means setting the threshold at $4\sigma$ for confidence.
\item The spin frequency of the star may change, and the star may have a measurable proper motion.  Both of these could cause the phase of the signal to drift with respect to a filter that does not take them into account.
\end{itemize}

Nevertheless, a suitably carefully designed analysis program could probably examine a single candidate in a one year data set in a minute or less on a teraflop computer.  This means that the known catalog of 800 pulsars will not present a problem for data analysis.

\section{Directed searches for accretion-driven radiation in binaries}
Although accreting systems may be promising sources of gravitational radiation, they also present a serious complication: the neutron star is always in orbit around a companion, and this induces a large intrinsic phase modulation in the signal.  This is in addition to the phase modulation induced by the motion of the detector, which can be removed (if the position of the star is known) by resampling for the solar barycenter as for known pulsars.  Not much attention has been paid to the filtering problem for an orbiting neutron star yet.  Given that the all-sky search for an isolated pulsar (below) is so demanding, it is safe to say that an all-sky search for an accreting source would be impossible in the near future.  But even for sources with known positions, there are major requirements:
\begin{itemize}
\item We must have lists of targets, so that we can search their positions.  Many targets will be {\em possible} accreting systems, such as nearby O/B stars that may harbor neutron stars in their envelopes.  More reliable are known X-ray binaries, provided the pulsar frequency as inferred from the X-ray observations is high enough. 
\item We know little about algorithms for searching over binary orbit parameters.  For known X-ray systems, the parameter space will be reasonably small, since the orbital period, eccentricity, and in some cases inclination will be known or constrained by X-ray observations.  But for neutron stars hidden inside O/B stars, the parameter space can be huge.  It is probably safe to assume the orbit is circular, but its inclination and period will not be known.
\end{itemize}

This is an urgent problem.  Detector groups will have to make an effort to develop lists of candidates and suitable algorithms.  The general outline of the algorithms will resemble the all-sky searches, to the extent that both kinds of analysis are searches of big parameter spaces.  But the details will be different, and in these details there may be possibilities for improving the efficiency of searches.

\section{All-sky survey}
This is the holy grail of gravitational pulsar astronomy.  We would like to discover sources down to the 1-year detector sensitivity limit over the whole sky.  As many people have stressed (see \cite{Brady}), this is impossible because the parameter space is too large.  This should not be surprising: no 
astronomical survey is ever done to the depth that directed searches can reach.  But progress is being made on understanding what is possible and how to achieve it.

Search algorithms can be classed as:
\begin{itemize}
\item {\bf Coherent}: They perform effectively a full matched-filter search for a given set of parameters over some observation time.  Signal-to-noise ratios increase as the square-root of observing time.  Parameter values must be chosen with a spacing that covers the parameter space with suitable sensitivity.  This issue is addressed by Krolak in his talk at this meeting, and was also considered in \cite{Brady}.  Because the size of the parameter space increases with the observation time, the available computer power sets a limit on the length of data that can be fully searched coherently.
\item {\bf Incoherent}: These methods combine short-period coherent searches in a way that loses phase information but retains some sensitivity to the continuing presence of a signal.  They typically require much less computer power than a coherent search over the same period of time, but they sacrifice sensitivity: signal-to-noise ratios increase as the fourth-root of the observing time.  The two incoherent approaches now under investigation are:
\begin{enumerate}
\item {\bf Stacking}, in which successive power spectra are added to enhance the signal.  This method is the more developed of the two, and its statistics are well understood.  See Brady's talk at this meeting for more details.
\item  {\bf Tracking}, in which a spectral line or feature is followed from one power spectrum to the next.  Less developed but in my opinion very promising (see Papa's talk at this meeting), this method offers similar statistics and sensitivity to stacking but works with essentially 1-bit data, so it has promise of improved efficiency.  
\end{enumerate}
\end{itemize}

These two methods will be combined into a search strategy: first perform a series of short-period coherent searches, then stack or track to identify a number of candidates over a longer period with a threshold lower than one will eventually set, then follow up the candidates with full-sensitivity coherent targeted searches (as for known pulsars) to eliminate those generated by noise.  The final step will still require parameter searches, but over smaller spaces.  The balance between coherent and incoherent searches, and the total sensitivity, will be set by the available computer power.

Brady will speak at this meeting about optimizing this strategy for a given amount of computer power and issue first addressed in this context by D.\ Nicholson .  I will illustrate the results of possible (non-optimized) strategies, but first I discuss a way that the {\em coherent} search itself may be done efficiently.  

\section{Efficient coherent search based on the Jones approximation}
Suppose we set our sights on doing a coherent search on data sets lasting one day.  Each search must still be performed for a large number of parameter sets, and it must be completed in a day in order to keep up with the data acquisition.  Such searches will inevitably involve Fourier transforms, and these are problematic on large data sets.  A data set lasting one day, sampled at 3~kHz to allow for a search up to 1.5~KHz (which means covering spin rates up to 750~Hz, probably not the fastest possible for neutron stars), would contain $2.6\E8$ samples, and would occupy 2~GB of memory if stored as double-precision floating point numbers.

In this section I suggest a method that may make searches in such long data sets more efficient.  This method is a variant and extension of one devised by G.\ S.\  Jones for his PhD thesis at Cardiff recently \cite{Jones}, so I will refer to it by his name.  It is related, I believe, to a method more recently used by Montero and Frasca (private communication) in the analysis of bar-detector data.  I will first describe how the method performs on a constant-frequency signal, compared with the standard Fourier transform, and then I will explain how I believe it can be developed to make efficient search algorithms for more complicated signals.  The method computes an {\em approximation} to the Fourier transform of the full data set, but the approximation is good enough to search for signals with a minimal loss of sensitivity.  

Most coherent search algorithms for the pulsar problem involve repeatedly performing FFT's.  In the simplest version of the matched-filtering method (see the talk by Krolak at this meeting), one performs the FFT of the data once and then, for each set of parameters, multiplies the data by the filter and performs an inverse Fourier transform.  A possible alternative, which I suggested some time ago \cite{SchutzBlair} and which was implemented by Jones \cite{Jones}, is to do the operation just described for widely spaced parameter values; for intermediate values one can ``step'' from one set to another with operations directly on the Fourier transform.  Both these methods will require an operation count that is at least of order 50--100 times the number of data points $N$ for the 1-day data set, in addition to the overhead of computing filters or other information for each set of parameters.  A different method, briefly described in \cite{Brady}, is to resample the data stream for each different parameter set in step with the phase of the expected signal, so that the signal can be found with a single FFT of the new samples.  Besides the costs of resampling, this method too requires 50--100~$N$ operations for each parameter set.  Methods that need to perform an FFT have additional overheads --- particularly communications costs --- when used on parallel computers.  It seems inevitable that 1-day searches will be performed on parallel computers, networks of computers, or both.

There are standard algorithms for performing FFT's on parallel machines with distributed memory, and they require only slightly more floating point operations than a single-processor FFT.  But they involve performing a matrix transpose within the parallel machine, a large exchange of data among all the processors.  This step can be time-consuming, especially in massively parallel machines or (even worse) when the computation involves machines distributed over a network.  

The Jones method suggested here avoids such a step.  For the accurate approximation of the Fourier transform of a {\em single} long data set of the size envisaged here, it requires about 1.6 times as many floating point operations as an FFT.  But if it is part of a search algorithm in which Fourier transforms are performed repeatedly, the Jones method can be much faster, essentially because it needs to perform part of its approximate Fourier transform only once.  This method is based essentially on a two-dimensional Fourier transform of the 1-dimensional data arrayed in a matrix.  It is highly parallelizable, and in fact has minimal communications overheads.

I assume that at the start of the analysis we will have a database of short-period Fourier transforms, perhaps the Fourier Transform Database that Astone discussed at this meeting.  The data sets are short enough that the signal has constant frequency within them, for any parameter set.  For our example data set, limited by the Doppler effect, these time-series sets last about 30 minutes, short enough that there is no measurable phase modulation up to 1.5~kHz.  So there will be something like 50 of these sets for a 1-day data run.  These 50 sets are the starting point for computing the approximation to the Fourier transform for each parameter set; the efficiency of the Jones method is in the fact that it does not have to perform the transform of each set repeatedly.  It happens only once (and even this is done as the data are acquired, before the 1-day analysis starts). 

To start the Jones-type analysis, one would read in the short-period FFT's and distribute different frequency ranges to different processors, so each processor would hold the 50 short-time FFT points for a range of frequencies $(f_0, \;f_1)$.  This band will be much wider than the spread of frequencies in any neutron-star signal.  All further processing involves only data in that set.  Therefore, the parallel machine does not even have to be a supercomputer: it could be several independent computers linked on a network, or even by the Internet!

I will illustrate the method by supposing that the data contains a signal of a {\em fixed} frequency during the whole period of 1 day.  I take the frequency for simplicity to be one that can be exactly resolved in the long-duration (1-day) transform, but it may not be resolved in the short transforms.   I will explain below how I believe the method can be generalised to perform matched filtering for any signal that does not stray outside the bandwidth of data held on a single processor.

Suppose that the  full 1-day time-series data set is ${x_j, \; j=0,...,n-1}$, with $n\sim2.6\E8$.  Split it up into $M\sim50$ short-period sets of length $N=n/M\sim5.2\E6$.  I will fix my notation by using Roman letters at the beginning of the alphabet as indices for the short-period sets, and Greek letters as labels for the sets themselves.  Thus, I can write the time-series data as
\begin{equation}
x_j = x_{a+N\alpha}, \quad\mbox{$a=0, ..., N-1$, and $\alpha = 0, ..., M-1$.}
\end{equation}

Let us see what happens to our assumed constant-frequency signal when a short-period FFT is performed.  The signal is a data set with a frequency index $k$, which can also be any integer in the range $(0,...,n-1)$.  Split it up into a set component index $\beta$ and a short-term frequency index $b$ (both integers because I have assumed the signal frequency can be resolved by the one-day transform) in the fashion
\begin{equation}
j = \beta + Mb, 
\end{equation}
so that the signal has the form
\begin{eqnarray}
x_j &=& \exp(2\pi ijk/n) \Rightarrow x_{a+N\alpha} = \exp\left[2\pi i(a+N\alpha)(\beta+Mb)/NM\right],\\
&=&\exp(2\pi ia\beta/NM)\exp(2\pi i\alpha\beta/M)\exp(2\pi iab/N).
\end{eqnarray}

The short-period FFT turns the signal into 
\begin{eqnarray}
\hat{x}_{b', \alpha} &=& \sum_a x_{a+N\alpha}\exp(-2\pi iab'/N),\\
&=& N\exp(2\pi i\alpha\beta/M)P_{\beta}(b'-b),
\end{eqnarray}
where I have defined 
\begin{equation}
P_\beta(b'-b) = \frac{1}{N}\sum_a\exp(2\pi ia\beta/NM)\exp[-2\pi ia(b-b')/N].
\end{equation}

This is quite complicated and will generally not stand out above noise in the short-period transforms.  But it becomes much clearer if we now transform on the set index, that is using data held only on a single processor.  Denoting the set transform by an overbar, we find
\begin{eqnarray}
\bar{\hat{x}}_{b',\beta'} &=& \sum_\alpha \hat{x}_{b',\alpha}\exp(-2\pi i\alpha\beta'/M),\\
&=& NP_\beta(b'-b)\sum_\alpha\exp[2\pi i\alpha(\beta-\beta')/M],\\
&=&nP_\beta(b-b')\delta_{\beta,\beta'}.
\end{eqnarray}

Before analysing this result, we should note that we have performed simply a 2-dimensional Fourier transform of the data matrix whose indices are $(a,\;\alpha)$.  It is easy to show that the operation count is {\em identical} to that of a single one-dimensional FFT of the original time-series.  

Now we look at the expression for $\bar{\hat{x}}_{b',\beta'}$.  It is not the Fourier transform, but it has only a single non-zero set element at each short-set frequency $b$, namely at index $\beta'$ equal to the inherent frequency set-index of the signal $\beta$.  So the effectiveness of Jones' method depends on how strongly peaked the function $P_\beta( c)$ is near $c=0$: if it concentrates the power there, then it will make the signal as visible as  in a normal Fourier transform.  This function is easy to calculate.  Its complex magnitude is essentially the central part of a sinc function:
\begin{equation}
|P_\beta(0)| = \sin(\pi\beta/M)/\sin(\pi\beta/n)/N\approx \frac{\sin(\pi\beta/M)}{\pi\beta/M}.
\end{equation}
For $\beta=0$ this is exactly 1, and in fact the Jones approximation exactly equals the  Fourier transform.  When $\beta>0$, the function $P_\beta(0)$ falls off gently, reaching 0.9 at $\beta=M/4$.  So the signal has 90\% of its amplitude over about one-quarter of the range of $\beta$.  It also has little power at other values of $\beta$: the ratio $P_\beta(1)/P_\beta(0)$ is no larger than 0.2 up to $\beta=M/4$.  For this reason, this approximate transform will pick out the signal essentially as well as a true Fourier transform over one-quarter of the range of $\beta$.

The whole range of $\beta$ can be covered in 4 steps by shifting on $\beta$ before performing the second Fourier transform.  This will be efficient, since the data for all $\beta$ are held on one processor.  These extra transforms require only of order $n\log_2(M)$ operations, since the first transform does not need to be done again.  The total operation count is then just 1.6 times that of a full FFT for the data set parameters I have taken here. 

On a closely coupled parallel computer this may or may not save time over alternative methods that involve matrix transposes.  If one is computing over a loosely coupled network (as may be necessary to get one Tflop of computing power), this might be faster than any kind of FFT.   But the main gain of this method is on repeated searches.  This is because the first part of the FFT, the short-set transform, is performed only once, so that the operation count for each further computation (allowing for 4 set-index transforms) is reduced by $4*\log_2(50)/log_2(2.6\E8)=0.8$.  This algorithm produces a good search with a speedup in operation count of a factor of 1.25, not counting savings on communications in the FFT algorithm and other potentially important filter-related savings discussed below.

Now I discuss the effect of the idealised assumptions I have made about the signal.  If the signal were not exactly resolvable in the long-term data set, then one could represent that by a value of $\beta$ that is not an integer.  This would not change the behaviour of the function $P_\beta$, but it would mean that one would expect the power to be distributed mainly between the two closest integers to $\beta$.  This is just what would happen in this case for a full FFT as well.  

To look for a more complicated signal, one would use the Jones method in a matched-filtering scenario.  Here one expects to multiply a Fourier transform by the filter and invert.  For our approximation, this might seem problematic, since we save operations only because we do not do the inverse transform of the short-period data.  If the filter required us to multiply the data by a function of the short-period frequency index $b$, our method would fail.  But I would argue that this is not necessary.  The short data sets are chosen to be of such a length that the signal does not change frequency during that period.  This means that the only effect of a filter on a short-index transform is to multiply it by a phase that is constant in $b$ and is only a function of $\beta$.  In this picture, one multiplies the short-period transforms by this $\beta$-dependent phase before performing the $\beta$-FFT, and the result should be as good as for the constant-frequency signal computed above.  

If this is right --- and these ideas need to be tested --- then there is an additional significant saving in the construction of filters for the signals we are searching for.  We need only compute $M$ complex numbers instead of $n$, to characterise the signal.  Essentially, we only need to compute the {\em average} phase correction for a signal over the duration of a single short-time data set.  This would make the generation of filters much more efficient, and that is a very significant overhead cost that has so far not been taken into account in estimation of the speed of search algorithms, to my knowledge.

It therefore seems to me that the Jones algorithm is well worth further study.

\acknowledgements{I wish to thank the following colleagues for useful discussions on the points covered here: G.S.\ Jones, M.-A.\ Papa, C.\ Cutler, K.\ Thorne, P.\ Brady, T.\ Creighton, and A.\ Krolak.}

\vfill 
\end{document}